\documentclass[aps,prb,reprint,amsmath,amsfonts,amssymb,showkeys]{revtex4-1}
\usepackage{graphicx,dcolumn,bm,color,microtype,hyperref}
\usepackage[version=4]{mhchem}

\newcommand{\br}{\bm{r}}
\newcommand{\rs}{r_s}
\newcommand{\eps}{\varepsilon}
\newcommand{\rsSB}{r_s^\text{SB}}
\newcommand{\mc}{\multicolumn}
\newcommand{\EHF}{e_\text{HF}^\text{FF}}
\newcommand{\tHF}{t_\text{HF}^\text{FF}}
\newcommand{\vHF}{v_\text{HF}^\text{FF}}
\newcommand{\ESBHF}{e_\text{HF}^\text{SB}}
\newcommand{\tSBHF}{t_\text{HF}^\text{SB}}
\newcommand{\vSBHF}{v_\text{HF}^\text{SB}}
\newcommand{\Ec}{e_\text{c}^\text{FF}}
\newcommand{\EcSB}{e_\text{c}^\text{SB}}

\newcommand{\Req}{R_\text{eq}}
\newcommand{\Rts}{R_\text{ts}}
\newcommand{\Edisso}{E_\text{disso}}
\newcommand{\Ets}{\Delta E_\text{ts}}
\newcommand{\alert}[1]{\textcolor{black}{#1}}

\begin{document}

\title{Symmetry-broken local-density approximation for one-dimensional systems}

\author{Fergus J. M. Rogers}
\affiliation{Research School of Chemistry, Australian National University, Canberra ACT 2601, Australia}
\author{Caleb J. Ball}
\email{caleb.ball@anu.edu.au}
\affiliation{Research School of Chemistry, Australian National University, Canberra ACT 2601, Australia}
\author{Pierre-Fran{\c c}ois Loos}
\email{pf.loos@anu.edu.au}
\thanks{Corresponding author}
\affiliation{Research School of Chemistry, Australian National University, Canberra ACT 2601, Australia}

\begin{abstract}
Within density-functional theory, the local-density approximation (LDA) correlation functional is typically built by fitting the difference between the near-exact and Hartree-Fock (HF) energies of the uniform electron gas (UEG), together with analytic perturbative results from the high- and low-density regimes. 
Near-exact energies are obtained by performing accurate diffusion Monte Carlo calculations, while HF energies are usually assumed to be the Fermi fluid HF energy. 
However, it has been known since the seminal work of Overhauser that one can obtain lower, symmetry-broken (SB) HF energies at any density. 
Here, we have computed the SBHF energies of the one-dimensional UEG and constructed a SB version of the LDA (SBLDA) from the results.
We compare the performance of the LDA and SBLDA functionals when applied to one-dimensional systems, including atoms and molecules. 
Generalization to higher dimensions is also discussed.
\end{abstract}


\maketitle

\section{
\label{sec:intro}
Introduction}
In 1965, Kohn and Sham \cite{Kohn65} showed that the knowledge of an analytical parametrization of the uniform electron gas (UEG) correlation energy \cite{WIREs16} allows one to perform approximate calculations for atoms, molecules and solids. \cite{ParrBook}
This led to the development of various local-density approximation (LDA) correlation functionals (VWN, \cite{Vosko80} PZ, \cite{Perdew81} PW92, \cite{Perdew92} \textit{etc.}), all of which require information on the high- and low-density regimes of the UEG, \cite{Ringium13, 1DEG13, gLDA14, Wirium14, Zia73, Isihara77, Rajagopal77, Glasser77, Isihara80, Glasser84, Seidl04, Chesi07, 2DEG11, Macke50, Bohm53, Pines53, GellMann57, DuBois59, Carr64, Misawa65, Onsager66, Wang91, Hoffman92, Endo99, Ziesche05, Sun10, 3DEG11, Handler12, Glomium11} and are parametrized using results from near-exact diffusion Monte Carlo (DMC) calculations. \cite{Lee11a, Tanatar89, Kwon93, Rapisarda96, Attaccalite02, Attaccalite03, GoriGiorgi03, Drummond09, Ceperley80, Ballone92, Ortiz94, Ortiz97, Kwon98, Ortiz99, Zong02, Drummond04, Spink13}

The LDA is the simplest approximation within density-functional theory (DFT). \cite{ParrBook}
It assumes that a real, non-uniform system (such as a molecule or a solid) may be treated as a collection of infinitesimally-small UEGs of electronic density $\rho$.
In principle, if one knows the reduced (i.e. per electron) correlation energy $\Ec(\rho)$ of the UEG for any density $\rho$, by summing the individual contributions over all space, it is therefore possible to obtain the LDA correlation energy
\begin{equation}        
\label{eq:EcLDA}
        E_\text{c}^\text{LDA} = \int \rho(\br) \,\Ec[\rho(\br)] \,d\br.
\end{equation}

Although it describes molecular bonding reasonably well compared to the Thomas-Fermi \alert{(TF)} model \cite{Thomas27, Fermi27} \alert{(which approximates the kinetic energy using the TF functional and ignores the exchange interaction between same-spin electrons)}, this rather crude approximation has managed only mixed success. \cite{Tong66}
In particular, the functional consistently overestimates correlation energies, giving rise to errors up to some factor of two. \cite{ParrBook}
Fortunately, errors in the exchange and correlation energies approximately compensate each other, thus generating a total energy that is usually in good agreement with experimental results. \cite{Ernzerhof97}

Since the emergent days of DFT, the UEG correlation energy $\Ec(\rho)$ has always been defined as the difference between the exact energy $e(\rho)$ and the Fermi fluid Hartree-Fock (FFHF) energy $\EHF(\rho)$.
However, in the early sixties, Overhauser \cite{Overhauser59, Overhauser62} showed that the FF state is never the Hartree-Fock (HF) ground state due to spin- and charge-density instabilities. \cite{VignaleBook}
Therefore, for any density, one can find a symmetry-broken HF (SBHF) solution which has a lower energy than the FFHF solution.
\alert{Unfortunately, the exact character of this SBHF solution is not given, nor is it necessary for it to remain the same over the full density range.}

Zhang and Ceperley \cite{Zhang08} have recently presented a computational ``proof'' of this statement. Performing unrestricted HF (UHF) calculations on the paramagnetic state of finite-size three-dimensional UEGs, they have succeeded in finding a ground state with broken spin-symmetry in the high-density region.
For lower densities, Trail et al.~discovered that the Wigner crystal (WC) is more stable than the FF state for $\rs > 1.44$ in 2D and $\rs > 4.5$ in 3D, \cite{Trail03} \alert{where the Wigner-Seitz radius $\rs$ is the average distance between electrons.}
These calculations were recently refined by Holzmann and coworkers. \cite{Delyon08, Bernu08, Bernu11, Baguet13, Baguet14, Delyon15}

\alert{Physically, a WC represents a state whose energy is dominated by the potential energy term, resulting in the electrons localizing on lattice points.}
\alert{This situation typically occurs at low densities, where the WC becomes the ground state.}
\alert{At high densities the kinetic energy dominates and the delocalized FF is the ground state.}
In addition to the usual FF and WC phases, they have also considered incommensurate crystals (a state in which the number of the charge density maxima is higher than the number of electrons), showing that such a phase is always favored over the FF; independently of the imposed polarization and crystal symmetry and in agreement with the earlier prediction of Overhauser. \cite{Overhauser59, Overhauser62}

Here, we propose to construct a symmetry-broken version of the LDA, taking the one-dimensional (1D) UEG as an example, and using SBHF energies instead of the usual FFHF expression.
From an experimental point of view, 1D systems have recently attracted much attention due to their practical realization in carbon nanotubes, \cite{SaitoBook, Egger98, Bockrath99, Ishii03, Shiraishi03} organic conductors, \cite{Schwartz98, Vescoli00, Lorenz02, Dressel05, Ito05} transition metal oxides, \cite{Hu02} edge states in quantum Hall liquids, \cite{Milliken96, Mandal01, Chang03} semiconductor heterostructures, \cite{Goni91, Auslaender00, ZaitsevZotov00, Liu05, Steinberg06} confined atomic gases, \cite{Monien98, Recati03, Moritz05} and atomic or semiconducting nanowires. \cite{Schafer08, Huang01}

This article is organized as follows. 
In Sec.~\ref{sec:1DEG}, we introduce the paradigm we have used to create a strict 1D UEG.
Section \ref{sec:SBHF} covers the acquisition of accurate SBHF energies at various densities, followed by the definition of the two correlation functionals in Sec.~\ref{sec:functional}.
These are compared in Sec.~\ref{sec:results} for 1D systems, including atoms and molecules.
Unless otherwise stated, atomic units are used throughout.

\begin{table*}
\caption{
\label{tab:HFenergy}
$\Delta \ESBHF$ (in millihartree) for various $n$ and $\rs$.
$\rsSB$ is the lowest value of $\rs$ for which one can find a SBHF solution.
}
\begin{ruledtabular}
\begin{tabular}{cccccccccccc}
			&			&	\mc{9}{c}{Wigner-Seitz radius $\rs=1/(2\rho)$}									\\
						\cline{3-11}																	
$n$			&	$\rsSB$	&	0.5		&	1		&	2		&	5		&	10		&	15		&	20		&	50		&	75		&	100		\\
\hline	
	9		&	1.22		&	$0$		&	$0$		&	$2.685$	&	$6.852$	&	$6.226$	&	$5.187$	&	$4.392$	&	$2.267$	&	$1.628$	&	$1.273$	\\
	19		&	0.79		&	$0$		&	$0.666$	&	$5.042$	&	$7.695$	&	$6.621$	&	$5.445$	&	$4.583$	&	$2.346$	&	$1.678$	&	$1.311$	\\
	29		&	0.64		&	$0$		&	$1.576$	&	$5.525$	&	$7.860$	&	$6.698$	&	$5.496$	&	$4.621$	&	$2.361$	&	$1.688$	&	$1.319$	\\
	39		&	0.55		&	$0$		&	$1.985$	&	$5.700$	&	$7.920$	&	$6.727$	&	$5.515$	&	$4.635$	&	$2.366$	&	$1.692$	&	$1.321$	\\
	49		&	0.50		&	$0$		&	$2.188$	&	$5.784$	&	$7.949$	&	$6.741$	&	$5.524$	&	$4.642$	&	$2.369$	&	$1.694$	&	$1.323$	\\	
	59		&	0.46		&	$0.000$	&	$2.302$	&	$5.830$	&	$7.965$	&	$6.749$	&	$5.529$	&	$4.646$	&	$2.370$	&	$1.695$	&	$1.324$	\\	
	69		&	0.43		&	$0.008$	&	$2.371$	&	$5.859$	&	$7.975$	&	$6.754$	&	$5.532$	&	$4.648$	&	$2.371$	&	$1.695$	&	$1.324$	\\	
	79		&	0.41		&	$0.143$	&	$2.418$	&	$5.878$	&	$7.982$	&	$6.757$	&	$5.534$	&	$4.650$	&	$2.372$	&	$1.695$	&	---		\\	
	89		&	0.40		&	$0.198$	&	$2.450$	&	$5.891$	&	$7.986$	&	$6.759$	&	$5.535$	&	$4.651$	&	$2.372$	&	---		&	---		\\	
	99		&	0.38		&	$0.244$	&	$2.473$	&	$5.901$	&	$7.989$	&	$6.760$	&	$5.536$	&	$4.652$	&	---		&	---		&	---		\\	
	109		&	0.36		&	$0.281$	&	$2.490$	&	---		&	---		&	---		&	---		&	---		&	---		&	---		&	---		\\	
	119		&	0.35		&	$0.312$	&	$2.503$	&	---		&	---		&	---		&	---		&	---		&	---		&	---		&	---		\\	
	129		&	0.35		&	$0.337$	&	---		&	---		&	---		&	---		&	---		&	---		&	---		&	---		&	---		\\	
	$\vdots$	&	$\vdots$	&	$\vdots$	&	$\vdots$	&	$\vdots$	&	$\vdots$	&	$\vdots$	&	$\vdots$	&	$\vdots$	&	$\vdots$	&	$\vdots$	&	$\vdots$	\\	
	$\infty$	&	0		&	$0.476$	&	$2.570$	&	$5.938$	&	$8.002$	&	$6.767$	&	$5.540$	&	$4.655$	&	$2.372$	&	$1.695$	&	$1.324$	\\	
\end{tabular}
\end{ruledtabular}
\end{table*}

\section{
\label{sec:1DEG}
One-dimensional uniform electron gas}
A 1D UEG is constructed by confining a number $n$ of interacting electrons to a ring of radius $R$ of electronic density \cite{QR12, Ringium13, gLDA14, NatRing16}
\begin{equation}
\label{eq:rho}
	\rho = \frac{n}{2 \pi R}=\frac{1}{2r_s}.
\end{equation}
where $\rs$ is the so-called Wigner-Seitz radius. \cite{VignaleBook, ParrBook} 
We refer the readers to Ref.~\onlinecite{Ringium13} for more details about this paradigm, which has been shown to be equivalent to the more conventional ``electrons-in-a-periodic-box'' model in the thermodynamic limit (i.e. $n \to \infty$), \cite{Glomium11, Ringium13, 1DEG13, Wirium14} but mathematically simpler. \cite{Glomium11}
This can be qualitatively explained by the ``short-sightedness'' of the electronic matter. \cite{Kohn96, Prodan05}
Because the paramagnetic and ferromagnetic states are degenerate in strict 1D systems, we will consider only the spin-polarized electron gas. \cite{Astrakharchik11, Lee11a, QR12, Ringium13, 1DEG13, gLDA14, Wirium14, 1DChem15} 

\begin{figure}
	\includegraphics[width=0.45\textwidth]{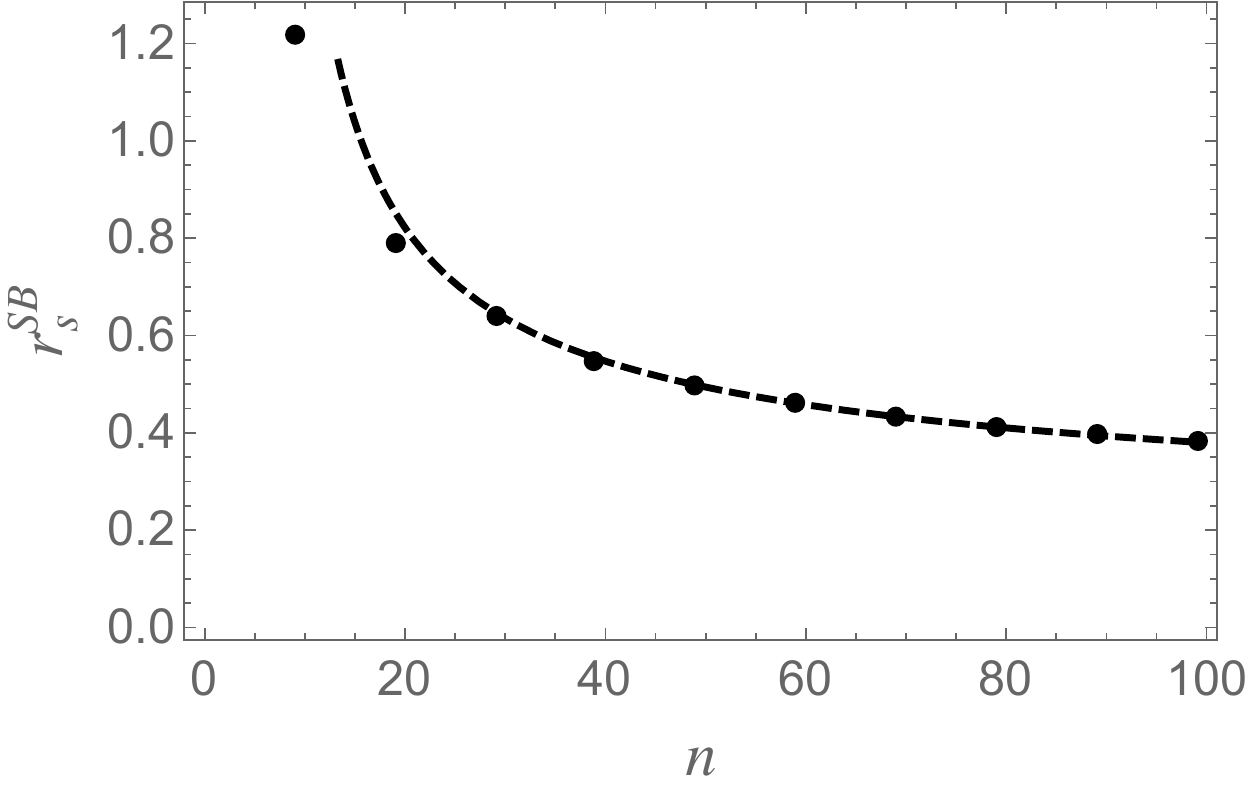}
\caption{
\label{fig:rsSB}
$\rsSB$ as a function of $n$.
The dashed curve represents the fit defined by Eq.~\eqref{eq:fit-rsSB}}
\end{figure}

The expression of the (non-symmetry-broken) FFHF energy is \cite{Ringium13}
\begin{equation}
\label{eq:eHF-1D}
	\EHF(\rs,n) = \frac{t_\text{HF}(n)}{\rs^2} + \frac{v_\text{HF}(n)}{\rs},
\end{equation}
with
\begin{subequations}
\begin{align}
	t_\text{HF}(n) & = \frac{\pi^2}{24} \frac{n^2-1}{n^2},
	\\
	v_\text{HF}(n)  & =  \left( \frac{1}{2} - \frac{1}{8n^2} \right) \left[ \psi\left(n+\frac{1}{2}\right) - \psi\left(\frac{1}{2}\right) \right] - \frac{1}{4},
\end{align}
\end{subequations}
where $\psi(x)$ is the digamma function. \cite{NISTbook}
This corresponds to occupying the $n$ lowest plane waves
\begin{equation}
\label{eq:PW}
	\phi_m(\theta) = \frac{\exp \left( i m\theta \right)}{\sqrt{2\pi R}},
\end{equation}
where $\theta$ is the angle of the electron around the ring and $m \in \mathbb{Z}$.
The FFHF energy \eqref{eq:eHF-1D} diverges logarithmically for large numbers of electrons, \cite{Schulz93, Fogler05a, Fogler05b, Ringium13}
\begin{equation}
	\EHF(\rs,n)  \sim \frac{\ln \sqrt{n}}{\rs} + O(1),
\end{equation}
but an identical divergence in the exact energy $e(\rs,n)$ results in a finite correlation energy
\begin{equation}
\label{eq:ec}
	\Ec(\rs,n) = e(\rs,n) - \EHF(\rs,n).
\end{equation}

As stated in Sec.~\ref{sec:intro}, our goal here is to determine the SB correlation energy in the thermodynamic limit
\begin{equation}
	\EcSB(\rs) = \lim_{n \to \infty} \EcSB(\rs,n)
\end{equation}
via extrapolation, where
\begin{equation}
\begin{split}
	\EcSB(\rs,n) 	& = e(\rs,n) - \ESBHF(\rs,n)  
	\\
				& = e(\rs,n) - \EHF(\rs,n)  + \Delta \ESBHF (\rs,n)
	\\
				& = \Ec(\rs,n) + \Delta \ESBHF (\rs,n).
\end{split}
\end{equation}

\begin{figure}
	\includegraphics[width=0.45\textwidth]{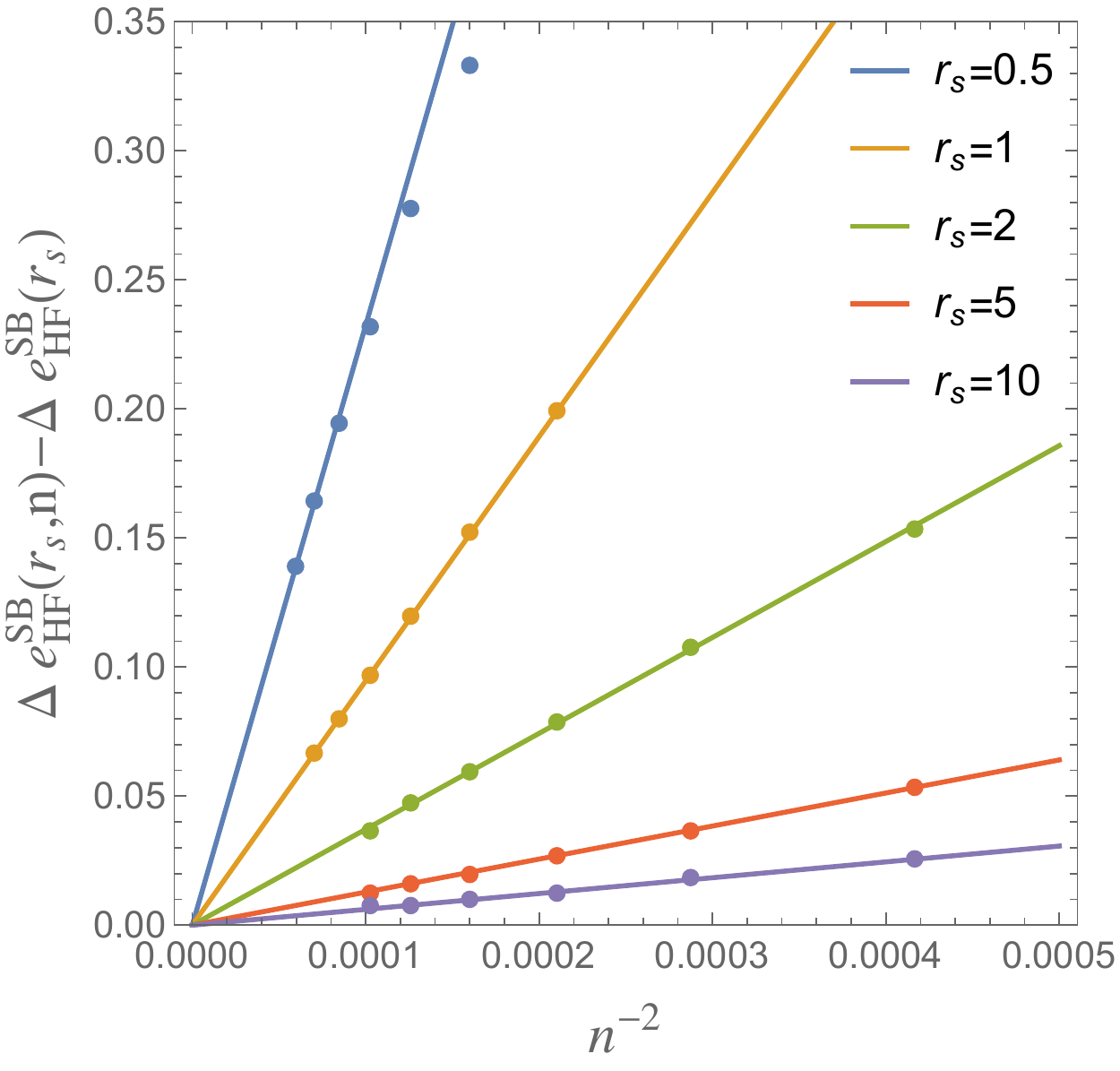}
\caption{
\label{fig:fit}
$\Delta \ESBHF (\rs,n)$ as a function of $n^{-2}$ for various $\rs$.
}
\end{figure}

\section{
\label{sec:SBHF}
Symmetry-broken Hartree-Fock calculations}
In order to obtain SBHF energies, we have written a self-consistent field program \cite{SzaboBook} using plane waves of the form \eqref{eq:PW} with 
\begin{equation}
	m = -\frac{M-1}{2}, \ldots, \frac{M-1}{2},
\end{equation}
where we have used up to $M = 399$ to ensure that our energies are always converged within microhartree accuracy.
As expected, large $\rs$ values require larger basis sets in order to converge the energy to the same accuracy due to the local character of the WC. \cite{TEOAS09}
The required one- and two-electron integrals can be found in Ref.~\onlinecite{Ringium13}.
The results are reported in Table \ref{tab:HFenergy}, where we have also reported $\rsSB$, the lowest $\rs$ value for which one can find a SBHF solution.
It is interesting to note that $\rsSB$ converges extremely slowly with respect to $n$, as shown in Fig.~\ref{fig:rsSB}.
We have found that the following \alert{function} (see Appendix \ref{app:model})
\begin{equation}
\label{eq:fit-rsSB}
	\rsSB(n) = \frac{a}{\ln n+b}
\end{equation}
(with $a = 1.13535$ and $b = -1.61346$) fits our data well. \footnote{We have used the $\rsSB$ values for $n = 49$ -- $99$ from Table \ref{tab:HFenergy} to obtain the values of $a$ and $b$ in Eq.~\eqref{eq:fit-rsSB}.}
Equation \eqref{eq:fit-rsSB} reveals that, in order to observe a SBHF solution below $\rs = 0.2$, one needs at least 1500 electrons.

We have obtained the thermodynamic values by extrapolation via the following asymptotic form:
\begin{equation}
\label{eq:extrap}
	\Delta \ESBHF (\rs,n) = \Delta \ESBHF (\rs) +A\,n^{-2}.
\end{equation}
For each $\rs$ value we have used the six largest $n$ values of Table \ref{tab:HFenergy}, except for $\rs = 0.5$ where only the last three terms were considered.
The quality of the fit is demonstrated in Fig.~\ref{fig:fit}.
A similar \alert{expression} to \eqref{eq:extrap} has been used by Lee and Drummond to extrapolate DMC calculations to the thermodynamic limit. \cite{Lee11a}

Physically, the appearance of the SBHF solution is characterized by the formation of a WC, i.e. where the electrons have ``crystallized'' such that they are separated by an angle of $2\pi/n$. \cite{Schulz93, Fogler05a, Fogler05b, Ringium13}
This phenomenon is easily understood in terms of the behaviour of the kinetic and potential energies with respect to $\rs$, which scale as $\rs^{-2}$ and $\rs^{-1}$ respectively (see Eq.~\eqref{eq:eHF-1D}).
Therefore, at small $\rs$, occupying the $n$ lowest plane waves is energetically favorable, as it minimizes the dominant kinetic energy contribution.
However, there exists a critical density $\rsSB$ for which it becomes compelling to break the spatial symmetry, (by populating higher plane waves) albeit at the expense of a meagre increase in the kinetic energy. 
This has the effect of lowering the potential energy by localizing the electrons, and in doing so, effectively reducing the interelectronic interaction. 

\begin{figure}
	\includegraphics[width=0.23\textwidth]{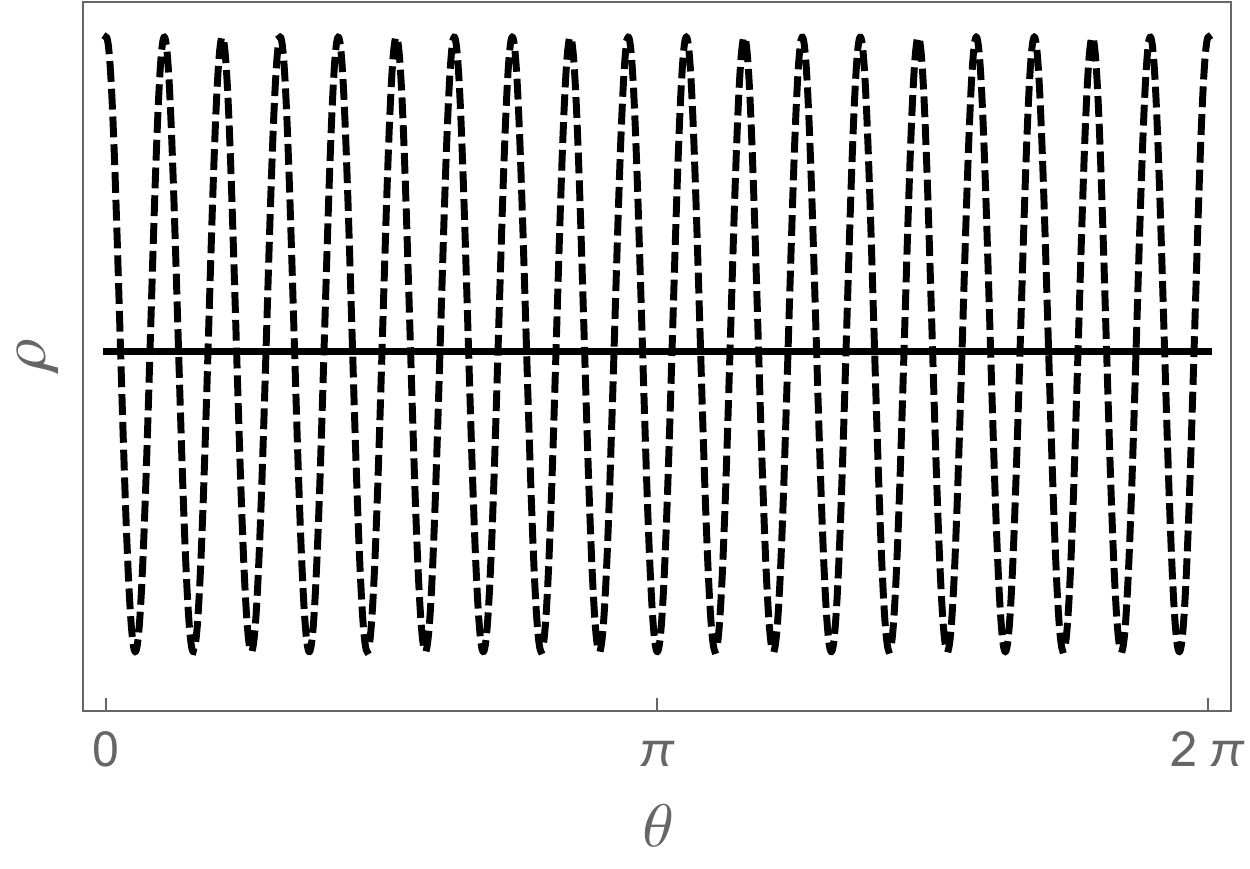}
	\includegraphics[width=0.23\textwidth]{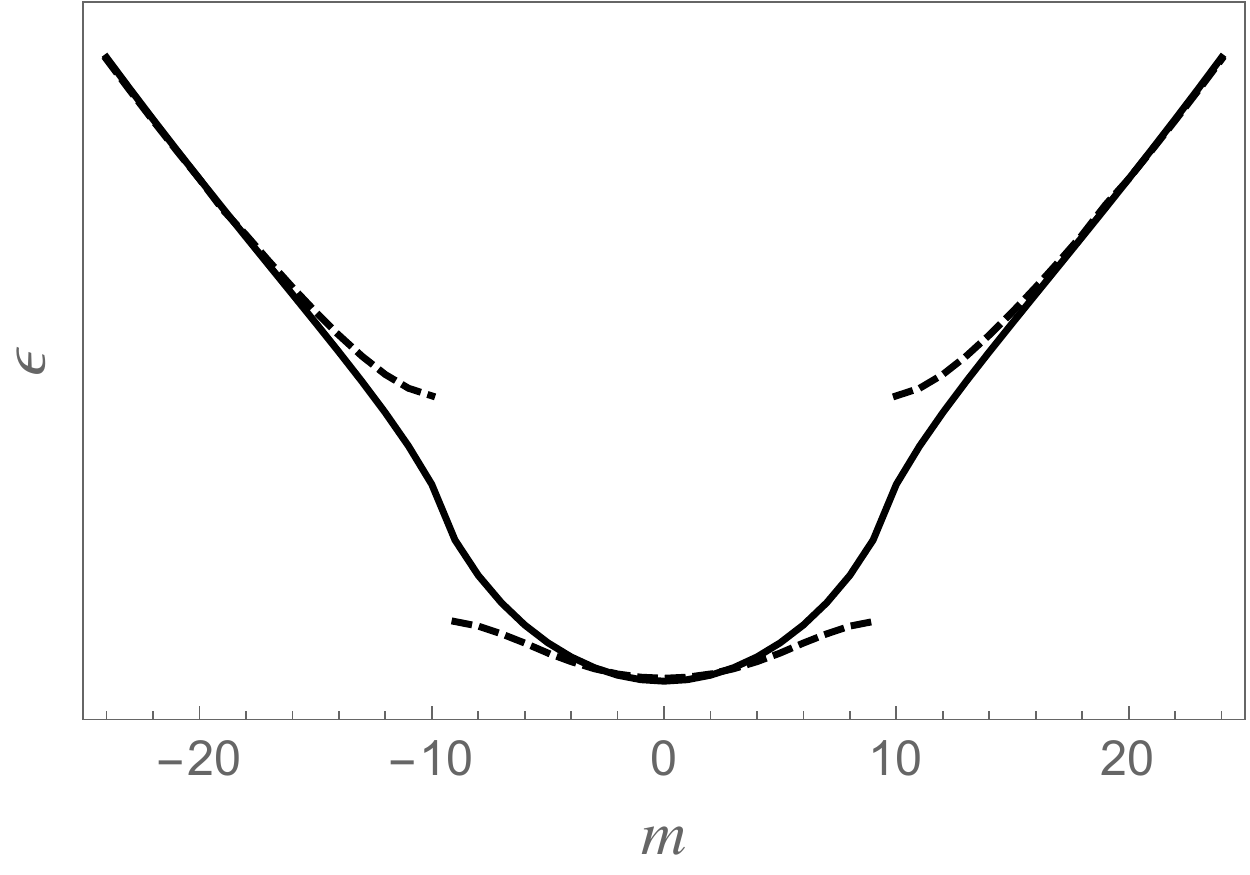}
\caption{
\label{fig:rho}
Electronic density $\rho$ as a function of $\theta$ (left) and orbital energies $\epsilon$ as a function of $m$ (right) for the HF (solid line) and SBHF (dashed line) solutions for $n=19$ at $\rs = 5$.
}
\end{figure}

The formation of a Wigner crystal is illustrated in Fig.~\ref{fig:rho} where we have represented the HF and SBHF densities for 19 electrons at $\rs = 5$.
One can see that, when the WC forms, a gap opens at the Fermi surface between the occupied and vacant orbitals.
As shown in Table \ref{tab:Ec}, the symmetry-breaking stabilization has a very significant effect on the values of the correlation energy, especially at intermediate and low densities where it represents a large fraction of the total correlation energy.
These results confirm Overhauser's prediction that in the thermodynamic limit, it is always favorable to break the spatial symmetry (see Appendix \ref{app:model}). 
We have, however, observed that the stabilization becomes extremely small at high density.
 
\begin{table}
\caption{
\label{tab:Ec}
Correlation energies (in millihartree) in the thermodynamic limit for various $\rs$.
}
\begin{ruledtabular}
\begin{tabular}{lccc}	
	$r_s$	&	$-\Ec$	&	$-\EcSB$			&	$1-\EcSB/\Ec$	\\
	\hline	
	0		&	$27.416$		&	$27.416$		&	0\%			\\
	0.5		&	$23.962$		&	$23.486$		&	2\%			\\
	1		&	$21.444$		&	$18.874$		&	12\%			\\
	2		& 	$17.922$		&	$11.984$		&	33\%			\\
	5		&	$12.318$		&	$4.316$		&	65\%			\\
	10		&	$8.292$		&	$1.525$		&	82\%			\\
	15		&	$6.319$		&	$0.779$		&	88\%			\\
	20		&	$5.133$		&	$0.478$		&	91\%			\\
	50		&	$2.476$		&	$0.104$		&	96\%			\\
	100		&	$1.358$		&	$0.034$		&	97\%			\\
\end{tabular}
\end{ruledtabular}
\end{table}

\section{
\label{sec:functional}
Correlation functionals}

\subsection{
\label{sec:LDA}
Local-density approximation}
In this study, we used the LDA functional developed in Ref.~\onlinecite{1DEG13}, which has been constructed using the ``robust'' interpolation proposed by Cioslowski \cite{Cioslowski12}
\begin{align}
\label{eq:Ec-LDA}
	e_\text{c}^\text{LDA}(\rs) & = t^2 \sum_{j=0}^{3} c_j t^j (1-t)^{3-j},		\\
	t & = \frac{\sqrt{1+4\,k\,\rs}-1}{2\,k\,\rs},
\end{align}
with
\begin{align}
	c_0 & = k\,\eta_0,	
	&
	c_1 & = 4\,k\,\eta_0+k^{3/2}\eta_1,
	\notag	\\
	c_2 & = 5\,\eps_0+\eps_1/k,
	&
	c_3 & = \eps_0,
	\notag
\end{align}
and the high- and low-density expansions, \cite{1DEG13} 
\begin{subequations}
\begin{align}
	e_\text{c}(\rs) = \eps_0 + \eps_1\,\rs + O(\rs^2),						\qquad \rs \ll 1,
	\\
	\label{eq:LDA-highrs}
	e_\text{c}(\rs) = \frac{\eta_0}{\rs} + \frac{\eta_1}{\rs^{3/2}},  + O(\rs^{-2}),	\qquad \rs \gg 1,
\end{align}
\end{subequations}
where
\begin{align}
	 \eps_0 & = -\frac{\pi^2}{360},
	 &
	 \eps_1 & = +0.00845,
	 \notag	\\
	 \eta_0 & = -\ln(\sqrt{2\pi})+3/4,
	 &
	 \eta_1 & = +0.359933	,
	 \notag
\end{align}
and $k=0.418268$ is a scaling factor, determined by a least-squares fit of the DMC data given in Refs.~\onlinecite{Lee11a} and \onlinecite{Ringium13}. 
As reported in Ref.~\onlinecite{1DEG13}, the LDA and DMC correlation energies agree to within 0.1 millihartree. 

\begin{figure}
	\includegraphics[width=0.45\textwidth]{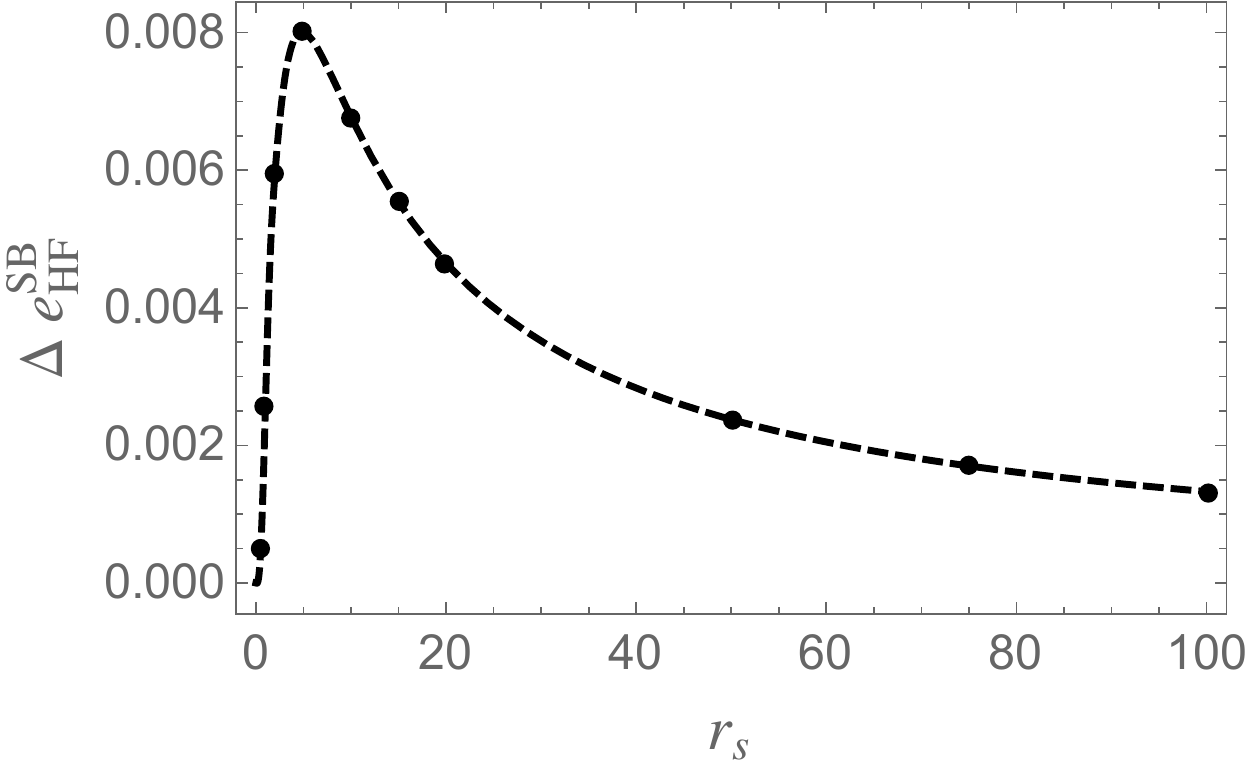}
\caption{
\label{fig:eSB}
$\Delta \ESBHF$ as a function of $\rs$ in the thermodynamic limit.
The dashed curve represents the fit defined by Eq.~\eqref{eq:fit-SBLDA}}
\end{figure}

\subsection{
\label{sec:SBLDA}
Symmetry-broken local-density approximation}
We define the SBLDA functional as
\begin{equation}
\label{eq:SBLDA-def}
	e_\text{c}^\text{SBLDA}(\rs) = e_\text{c}^\text{LDA}(\rs) + \Delta \ESBHF (\rs),
\end{equation}
where $e_\text{c}^\text{LDA}(\rs)$ is given by \eqref{eq:Ec-LDA}.
We propose to use the following \alert{expression} for the symmetry-breaking stabilization:
\begin{equation}
\label{eq:fit-SBLDA}
	\Delta \ESBHF (\rs) = \rs^2\frac{a_0 + a_1 \rs + a_2 \rs^2 - \eta_0 \rs^3}{b_0+ b_1 \rs^5 + b_2 \rs^{11/2} + \rs^6}
\end{equation}
where 
\begin{align}
	a_0 & =  -0.0646228,
	&
	a_1 & = 0.535062,
	&
	a_2 & = -0.490719,
	\notag	\\	
	b_0 & =53.1171,
	& 
	b_1 & = 1.53114,
	& 
	b_2 & = 2.19606.
	\notag
\end{align}
The expression \eqref{eq:fit-SBLDA} is illustrated in Fig.~\ref{fig:eSB} alongside the corresponding data of Table \ref{tab:Ec}.
The quality of \eqref{eq:fit-SBLDA} is excellent with a maximum error of 7 microhartrees compared to the values reported in Table \ref{tab:HFenergy}.

In the low-density limit, Eq.~\eqref{eq:fit-SBLDA} behaves as
\begin{equation}
	\Delta \ESBHF (\rs) = -\frac{\eta_0}{\rs} + O(\rs^{-3/2}),	\alert{\qquad \rs \gg 1,}
\end{equation}
\alert{
which, when combined with Eqs.~\eqref{eq:LDA-highrs} and \eqref{eq:SBLDA-def}, gives
\begin{equation}
	e_\text{c}^\text{SBLDA}(\rs) = O(\rs^{-3/2}),	\qquad \rs \gg 1.
\end{equation}
This behaviour is a direct consequence of the SBHF treatment allowing the electrons to localize at low densities, unlike the FFHF solution.
It is therefore able to correctly identify the appearance of the WC phase, which results in the low-density expansion of the SBHF solution matching that of the exact energy up to $O(\rs^{-3/2})$.
This difference in low-density behaviour creates an important distinction between $e_\text{c}^\text{LDA}$ and $e_\text{c}^\text{SBLDA}$.
}
For small $\rs$, we have imposed $\Delta \ESBHF (\rs)$ to be quadratic in $\rs$, as shown in Appendix \ref{app:model}.

\begin{table}
	\caption{\label{tab:IEsEAs} Ionization energy and electron affinity (in eV) of 1D atoms calculated with various methods.}
	\small
	\begin{ruledtabular}
		\begin{tabular}{lcccccc}
						&		\mc{3}{c}{Ionization energy}		&	\mc{3}{c}{Electron affinity}		\\
			Atom		&		\mc{3}{c}{\ce{A -> A+ + e-} }			&	\mc{3}{c}{\ce{A + e- -> A-}}		\\
									\cline{2-4}									\cline{5-7}
						&	MP3		&	LDA		&	SBLDA	&	MP3		&	LDA		&	SBLDA	\\
			\hline
			\ce{H}		&	13.606	&	14.125	&	14.013	&	3.893	&	4.327	&	4.154	\\
			\ce{He}		&	33.895	&	34.393	&	34.325	&	---		&	---		&	0		\\
			\ce{Li}		&	4.522	&	4.895	&	4.712	&	1.395	&	1.717	&	1.512	\\
			\ce{Be}		&	10.408	&	10.822	&	10.669	&	---		&	---		&	0		\\
			\ce{B}		&	2.099	&	2.386	&	2.190	&	0.638	&	0.875	&	0.688	\\
			\ce{C}		&	4.730	&	5.056	&	4.865	&	---		&	---		&	0		\\
			\ce{N}		&	1.14		&	1.38		&	1.20		&	0.34		&	0.51		&	0.37		\\
			\ce{O}		&	2.56		&	2.83		&	2.63		&	---		&	---		&	---		\\
			\ce{F}		&	0.68		&	0.87		&	0.72		&	0.2		&	0.3		&	0.2		\\
			\ce{Ne}		&	1.5		&	1.7		&	1.5		&	---		&	---		&	---		\\
		\end{tabular}		
	\end{ruledtabular}		
\end{table}		

\section{
\label{sec:results}
Correlation energy in one-dimensional systems}
In this Section, we test the LDA and SBLDA functionals defined in Secs.~\ref{sec:LDA} and \ref{sec:SBLDA}, respectively.
The LDA and SBLDA correlation energies are obtained via
\begin{equation}
	E_\text{c}^\text{LDA/SBLDA} = \int \rho(x) e_\text{c}^\text{LDA/SBLDA}[\rho(x)] dx.
\end{equation}
which is computed by numerical quadrature.
These quantities are calculated with the HF density, i.e. they are not calculated self-consistently. \cite{Pople92}
However, as expected, \cite{Johnson93} we have observed that the differences between self- and non-self-consistent densities are extremely small. \cite{gLDA14, Wirium14}
In some cases, we have also reported the exact, MP2 and MP3 energies. \cite{SzaboBook, Helgaker}
For additional information about these calculations, we refer the readers to Refs.~\onlinecite{gLDA14} and \onlinecite{1DChem15} where theoretical and computational details are provided.

\begin{table}
\caption{
\label{tab:TableMol}
Equilibrium bond length $\Req$, transition structure bond length $\Rts$, dissociation energy $\Edisso$ and transition barrier $\Ets$ of 1D molecules calculated with various methods.}
\begin{ruledtabular}
\begin{tabular}{llcccc}
Method		&				&	\mc{4}{c}{Molecules}																	\\
							\cline{3-6}
			&				&	\ce{H2+}		&	\ce{HeH^2+}	&	\ce{He2^3+}	&	\ce{H2}		\\
\hline
Exact		& 	$\Req$		&	2.581		&	2.182		&	1.793		&	2.639		\\
			& 	$\Rts$		&	---			&	3.296		&	4.630		&	---			\\
			& 	$\Edisso$		&	$-0.3307$		&	$+0.1697$	&	$+0.0131$	&	$-0.1859$	\\
			&	$\Ets$		&	---			&	$+0.0209$	&	$+0.2924$	&	----			\\
\hline
HF			& 	$\Req$		&	2.581		&	2.182		&	1.793		&	2.636		\\
			& 	$\Rts$		&	---			&	3.296		&	4.630		&	---			\\
			& 	$\Edisso$		&	$-0.3307$		&	$+0.1697$	&	$+0.0131$	&	$-0.1846$	\\
			&	$\Ets$		&	---			&	$+0.0209$	&	$+0.2924$	&	----			\\
\hline
MP2			& 	$\Req$		&	2.581		&	2.182		&	1.793		&	2.637		\\
			& 	$\Rts$		&	---			&	3.296		&	4.630		&	---			\\
			& 	$\Edisso$		&	$-0.3307$		&	$+0.1697$	&	$+0.0131$	&	$-0.1854$	\\
			&	$\Ets$		&	---			&	$+0.0209$	&	$+0.2924$	&	----			\\
\hline
MP3			& 	$\Req$		&	2.581		&	2.182		&	1.793		&	2.638		\\
			& 	$\Rts$		&	---			&	3.296		&	4.630		&	---			\\
			& 	$\Edisso$		&	$-0.3307$		&	$+0.1697$	&	$+0.0131$	&	$-0.1857$	\\
			&	$\Ets$		&	---			&	$+0.0209$	&	$+0.2924$	&	----			\\
\hline
LDA			& 	$\Req$		&	2.573		&	2.176		&	1.790		&	2.627		\\
			& 	$\Rts$		&	---			&	3.291		&	4.630		&	---			\\
			& 	$\Edisso$		&	$-0.3	328$		&	$+0.1696$	&	$+0.0126$	&	$-0.1857$	\\
			&	$\Ets$		&	---			&	$+0.0214$	&	$+0.2964$	&	----			\\
\hline
SBLDA		& 	$\Req$		&	2.564		&	2.172		&	1.788		&	2.619		\\
			& 	$\Rts$		&	---			&	3.285		&	4.619		&	---			\\
			& 	$\Edisso$		&	$-0.3341$		&	$+0.1697$	&	$+0.0124$	&	$-0.1864$		\\
			&	$\Ets$		&	---			&	$+0.0216$	&	$+0.2995$	&	----			\\
\end{tabular}		
\end{ruledtabular}
\end{table}	

\subsection{
\label{sec:box}
Two electrons in a box}
\begin{figure}
	\includegraphics[width=0.45\textwidth]{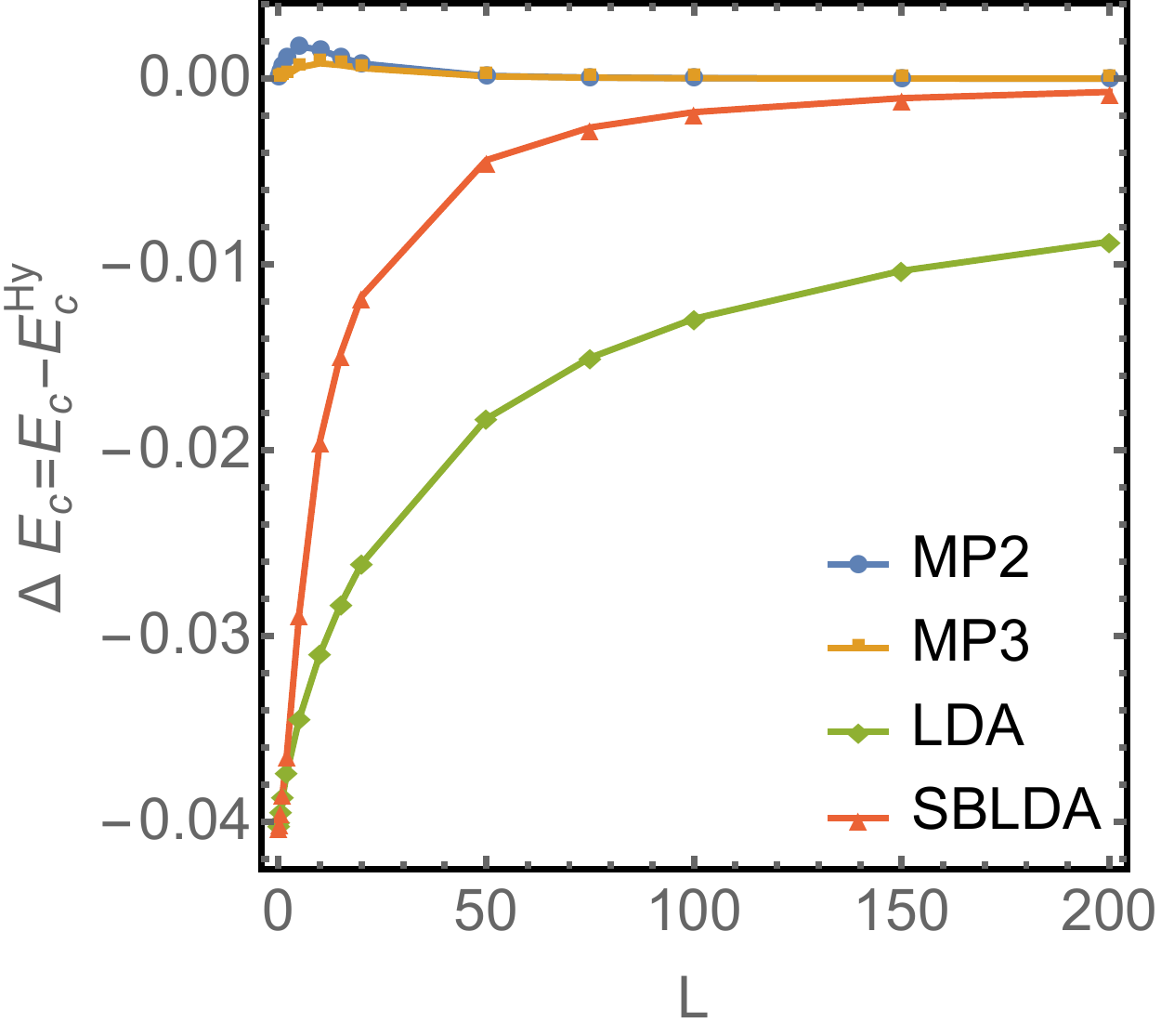}
\caption{
\label{fig:box}
Error in correlation energy $\Delta E_\text{c}$ for two electrons in a box of length $L$ with MP2, LDA and SBLDA.
}
\end{figure}

As a simple example to illustrate the performance of the SBLDA functional, we have thoroughly studied the well-known system composed by two electrons in a box of length $L$.
In particular, we report variations in the correlation energy as a function of $L$ obtained using the MP2, MP3, LDA and SBLDA methods.
In the interest of completion, we have also calculated the exact correlation energy of this system using a Hylleraas-type (Hy) expansion. \cite{Hylleraas28, Hylleraas29, Hylleraas30, Hylleraas64, Ballium10}
The results of this investigation are illustrated in Fig.~\ref{fig:box}, where the error in the correlation energy $\Delta E_\text{c} =E_\text{c} - E_\text{c}^\text{Hy}$ is resolved as a function of $L$.
As expected, MP2 and MP3 yields reliable estimates of the correlation energy for this system.
On the other hand, LDA returns a poor estimate of this energy. 
This discrepancy is especially clear in the high-density region (i.e. small  $L$).
By construction, LDA and SBLDA have similar performances at high densities.
However, SBLDA is much more accurate than LDA at low densities, where stabilization returned from breaking the symmetry is most significant.
Similar results are expected for different external potentials. \cite{EcLimit09, EcProof10, Frontiers10}

\subsection{
\label{sec:1Datoms}
Atoms}
We have calculated the ionization energies and electron affinities of 1D atoms \cite{1DChem15} using the \textsc{Chem1D} software developed by one of the authors. \cite{Ball15}
The values obtained with the LDA and SBLDA functionals are compared to the MP3 values in Table \ref{tab:IEsEAs}\alert{, which has been observed to be an exceptionally accurate method in such systems.\cite{1DChem15}}
Overall, LDA and SBLDA overestimate the ionization energies and electron affinities for these systems.
It is interesting to note that, although the performance of the LDA and SBLDA functionals are quite poor compared to \alert{MP3} for small atoms, the results become rapidly more accurate for larger atoms.
In particular, we observe that the accuracy of SBLDA improves faster than LDA.
For example, although the deviation between MP3 and SBLDA is only 0.04 eV for the ionization energy of the \ce{F} atom, the LDA is still 0.19 eV off.
This effect is most easily by acknowledging that larger atoms have more diffuse orbitals which possess lower density regions. \cite{1DChem15}

\subsection{
\label{sec:H2p}
One-electron diatomics}
\begin{figure}
	\includegraphics[width=0.45\textwidth]{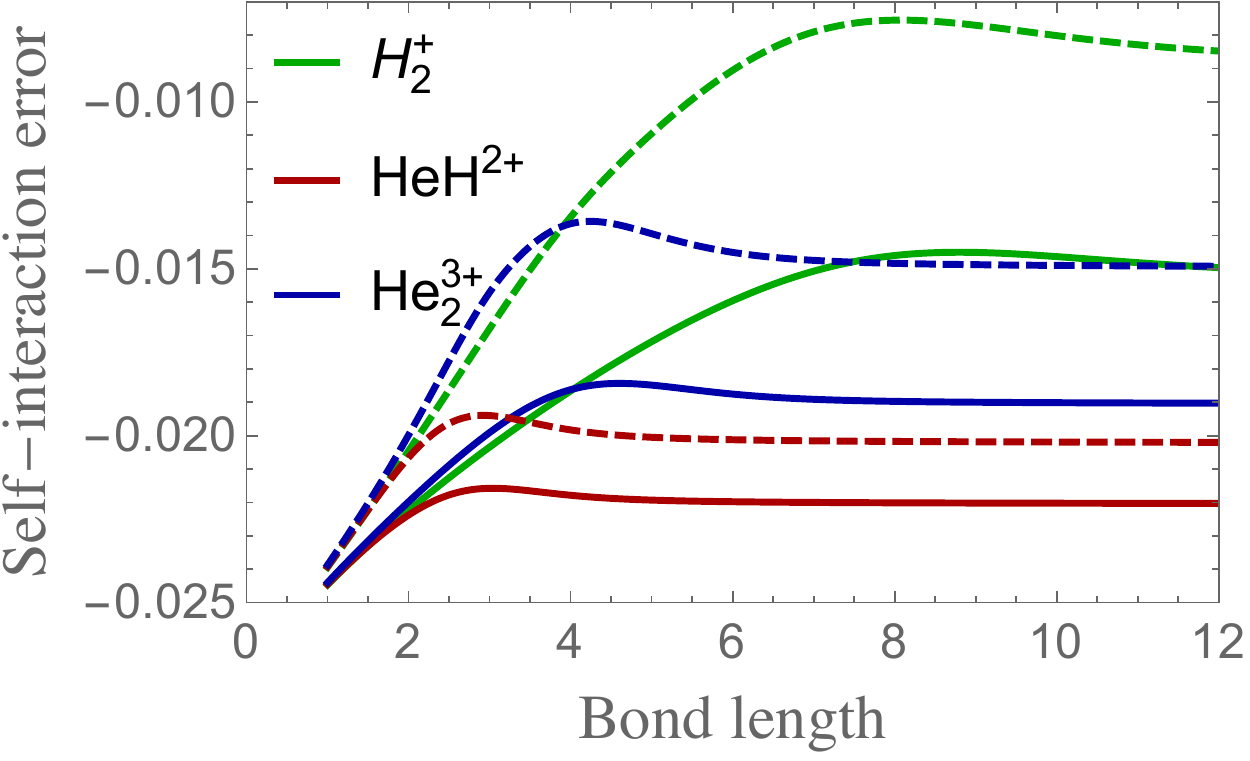}
\caption{
\label{fig:SIE}
Self-interaction error in \ce{H2+} (green), \ce{HeH^2+} (red) and \ce{He2^3+} (blue) calculated with LDA (solid) and SBLDA (dashed) as a function of the bond length.
}
\end{figure}

As commonly reported, LDA-type functionals suffer from the self-interaction error (SIE), \cite{Merkle92, Savin96, Perdew97, Zhang98, Tsuneda14} i.e. the unphysical interaction of an electron with itself.
This phenomenon is also known as the delocalization error and can be understood as the tendency of approximate functionals to artificially spread the electron density. \cite{Cohen08a, Cohen08b, MoriSanchez08, Cohen12}
In Fig.~\ref{fig:SIE}, we have reported the SIE in the one-electron diatomic molecules \ce{H2+}, \ce{HeH^2+} and \ce{He2^3+} as a function of the bond length.
Although the SIE is obviously still present in SBLDA, one can see that it is less pronounced than it is in LDA.
This statement is true at any bond length for the three molecules considered here.
As an illustration, we have computed the energy of the \ce{H} atom to be $-0.5$, $-0.519054$ and $-0.514971$ for HF, LDA and SBLDA, respectively.

As reported in Table \ref{tab:TableMol}, the dissociation energy of \ce{H2+} is $-0.3307$, $-0.3328$ and $-0.3341$ for HF, LDA and SBLDA, respectively.
Like in three dimensions, LDA and SBLDA slightly overestimate the binding energy of \ce{H2+}.
The equilibrium bond lengths are $2.581$, $2.573$ and $2.564$, showing that LDA and SBLDA predict bond lengths that are slightly too short.

As reported in Ref.~\onlinecite{1DChem15}, \ce{HeH^2+} and \ce{He2^3+} are metastable, and it is instructive to know if LDA and SBLDA can predict this peculiar feature properly.
Table \ref{tab:TableMol} reports the equilibrium bond length of these molecules as well as the transition structure bond lengths and the height of the barrier.
As we have observed in \ce{H2+}, the bond lengths predicted by LDA and SBLDA are slightly too short while the transition state barriers are overestimated.
It is interesting to note that both LDA and SBLDA predict (correctly) \ce{HeH^2+} and \ce{He2^3+} as being metastable species.

\subsection{
\label{sec:H2}
Dissociating H$_2$}
\begin{figure}
	\includegraphics[width=0.45\textwidth]{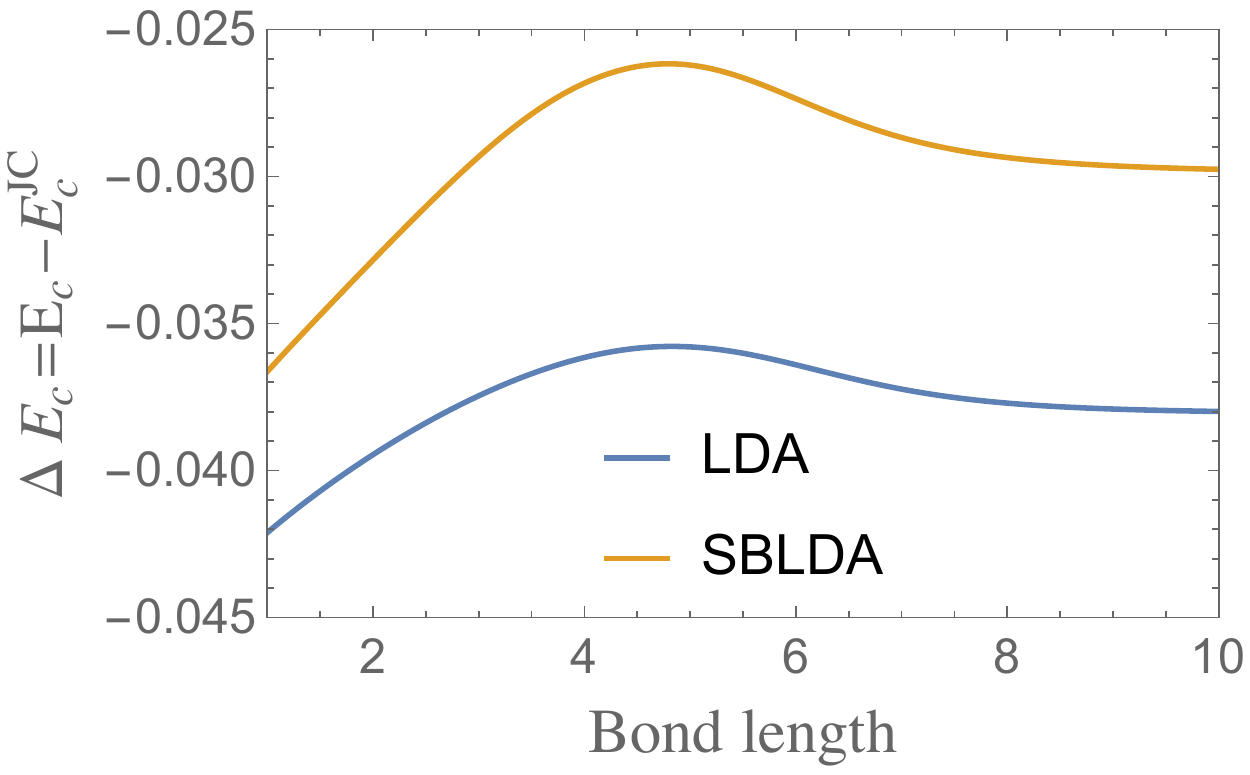}
\caption{
\label{fig:H2}
Error in correlation energy $\Delta E_\text{c}$ in \ce{H2} calculated with LDA and SBLDA as a function of the bond length.
}
\end{figure}

The apparently simple problem of stretching \ce{H2} has been widely studied within DFT, as it reveals a common pitfall of approximate density functionals known as the static correlation error. \cite{Cohen08a, Cohen08c, Cohen12}
Figure \ref{fig:H2} displays the error in the correlation energy ($\Delta E_\text{c}$) of \ce{H2}, as computed by the LDA and SBLDA functionals, as a function of the bond length.
The exact results have been obtained with a James-Coolidge-type (JC) expansion. \cite{James33, 1DChem15}
Although the error in the SBLDA correlation energy is still significant, we observe a clear improvement for all bond lengths compared to LDA.
This result is encouraging given the simplicity of the SBLDA functional.

Table \ref{tab:TableMol} reports the equilibrium bond length of \ce{H2} as well as its dissociation energy obtained with HF, MP2, MP3, LDA and SBLDA.
Compared to the exact results we observe, as previously reported, \cite{1DChem15} that MP2 and MP3 are extremely accurate in 1D.
In contrast to the observations of Sec.~\ref{sec:H2p}, the LDA slightly underestimates the dissociation energy here (agreeing with MP3), while the SBLDA continues to overestimate the same value. Both functionals, however, continue to yield a shorter bond length.

\section{Discussion and concluding remarks}
Inspired by Overhauser's forecasts made some fifty years ago, we have constructed a symmetry-broken (SB) version of the commonly-used local-density approximation (LDA) for one-dimensional systems.
The newly designed functional, which we have named SBLDA, has shown to surpass the performance of its LDA parent in providing better estimates of the correlation energy.
More importantly, we believe that this functional could be potentially useful as a superior starting point for more accurate approximations within density-functional theory (DFT), such as generalized gradient approximations (GGAs) or hybrid functionals. \cite{Becke14}
The methodology presented here is completely general and can be applied to higher-dimensional systems, where SB Hartree-Fock calculations have already been performed. \cite{Trail03, Bernu11, Baguet13, Baguet14}
The design of new exchange and correlation functionals for two- and three-dimensional systems based on the idea developed here is currently under progress in our group.

\begin{acknowledgments}
The authors would like to thank Peter Gill for many enlightening discussions about one-dimensional chemistry.
P.F.L.~thanks the Australian Research Council for a Discovery Early Career Researcher Award (DE130101441) and a Discovery Project grant (DP140104071).
P.F.L.~also thanks the NCI National Facility for generous grants of supercomputer time. 
C.J.B. is grateful for an Australian Postgraduate Award.
\end{acknowledgments}

\appendix

\section{
\label{app:model}
Behavior of $\Delta \ESBHF$ near $\rsSB$}

Near the critical density $\rsSB$, it is possible to use a simple two-orbital model to study the symmetry-breaking process.
When the symmetry breaking occurs, a small energy gap appears at the Fermi surface thanks to the mixing of the HOMO $\phi_{\pm \frac{n-1}{2}}(\theta)$ and LUMO $\phi_{\pm \frac{n+1}{2}}(\theta)$.
Therefore, to study the behavior of $\Delta \ESBHF$ near $\rsSB$ (see Sec.~\ref{sec:SBLDA}), we consider the two orthonormalized molecular orbitals 
\begin{subequations}
\begin{align}
	\varphi_1(\theta) & = \frac{\phi_{\frac{n-1}{2}}(\theta) + c\,\phi_{-\frac{n+1}{2}}(\theta)}{\sqrt{1+c^2}},
	\\
	\varphi_2(\theta) & = \frac{\phi_{-\frac{n-1}{2}}(\theta) + c\,\phi_{\frac{n+1}{2}}(\theta)}{\sqrt{1+c^2}}.
\end{align}
\end{subequations}
This two-orbital model teaches us that there exists a critical density
\begin{equation}
\label{eq:rsSB-HL}
	\rsSB(n) = \frac{\pi^2/4}{\psi(n-1/2)+\frac{1}{2n-1}+\gamma+2(\ln 2-1)}
\end{equation} 
(where $\gamma$ is the Euler-Mascheroni constant \cite{NISTbook}) after which it is energetically favourable to mix these two orbitals, and the value
\begin{equation}
	c^2 = \frac{1 - \rsSB/\rs}{1 + \rsSB/\rs + \frac{4}{\pi^2} (\frac{1}{2n+1}+\frac{1}{2n-1}) \rsSB}
\end{equation} 
minimizes the energy for $\rs > \rsSB$.
Before $\rsSB$, the solution is the usual FF state of energy
\begin{equation}
	\EHF	(\rs,n) = \frac{\tHF(n)}{\rs^2} + \frac{\vHF(n)}{\rs},
\end{equation}
where
\begin{subequations}
\begin{align}
	\tHF(n)	 & = \frac{\pi^2 (n-1)^2}{4n^2},
	\\
	\vHF(n) & = \frac{\psi(n-1/2)+\gamma+2\ln 2}{n} ,
\end{align}
\end{subequations}
minimizes the kinetic energy.
However, for $\rs > \rsSB$, this is outweighed by negative contributions in the potential term that drive the symmetry-breaking process.
The kinetic and potential parts of the symmetry-breaking stabilization are given by
\begin{subequations}
\begin{align}
	\ESBHF(\rs,n) & = \EHF	(\rs,n) - \Delta \ESBHF(\rs,n),
	\\
	\Delta \ESBHF(\rs,n) & = \Delta \tSBHF(\rs,n) + \Delta \vSBHF(\rs,n),
\end{align}
\end{subequations}
with
\begin{align}
	&\Delta \tSBHF(\rs,n)  = - \frac{\pi^2 c^2}{(1+c^2) n \rs^2},
	\\
	&\Delta \vSBHF(\rs,n) 		= \frac{4c^2}{(1+c^2)^2 n \rs}  	\notag
	\\
	&	\qquad \qquad \times \left[ \frac{\pi^2/4}{\rsSB(n)} - n c^2 \left( \frac{1}{2n-1} - \frac{1}{2n+1} \right) \right].
\end{align}
In the thermodynamic limit, Eq.~\eqref{eq:rsSB-HL} yields
\begin{equation}
	\rsSB \sim \frac{\pi^2/4}{\ln n+\gamma+2(\ln 2-1)},
\end{equation}
which has motivated our use of a similar expression in Eq.~\eqref{eq:fit-rsSB}. 
Because $\lim_{n \to \infty} \rsSB(n) = 0$, it also proves that, in the thermodynamic limit, there must exist a SBHF solution for any $\rs > 0$ in agreement with Overhauser's results. \cite{Overhauser59, Overhauser62}
Expanding $\Delta \ESBHF$ at $\rs \sim \rsSB$ yields
\begin{multline}
	\Delta \ESBHF(\rs,n) =
	\frac{\pi^2/4}{n\left[ \frac{\rsSB}{\pi^2/2} \left( \frac{1}{2n-1} + \frac{1}{2n+1} \right) + 1 \right]} \frac{\left( \rs -\rsSB \right)^2}{(\rsSB)^4} 
	\\ + O\left[\left( \rs -\rsSB \right)^3\right],
\end{multline}
showing that the behavior of $\Delta \ESBHF$ is quadratic near $\rsSB$.

\end{document}